\documentclass[useAMS,usenatbib,usegraphicx]{mn2e}

\usepackage{graphicx}
\usepackage[breaklinks,colorlinks,citecolor=black,linkcolor=black,urlcolor=black]{hyperref}

\newcommand*{\secref}{Section}

\newcommand*{\eqnref}{Equation}
\newcommand*{\figref}{Figure}
\newcommand*{\tabref}{Table}
\newcommand*{\transpose}[1]{\ensuremath{#1^{\mathrm{T}}}}
\newcommand*{\normal}{\ensuremath{\mathcal{N}}}
\newcommand*{\wishart}{\ensuremath{\mathcal{W}}}
\newcommand*{\one}{\ensuremath{\mathbf{1}}}
\newcommand*{\mysub}[2]{\ensuremath{#1_{\mathrm{#2}}}}
\newcommand*{\Mgas}{\mysub{M}{gas}}
\newcommand*{\ltsim}{\ {\raise-.75ex\hbox{$\buildrel<\over\sim$}}\ }

\makeatletter
\DeclareRobustCommand\mit{\@fontswitch{\relax}{\mathchoice}}
\def\setboxz@h{\setbox\z@\hbox}
\def\wdz@{\wd\z@}
\def\boxz@{\box\z@}
\def\setbox@ne{\setbox\@ne}
\def\wd@ne{\wd\@ne}
\def\math@atom#1#2{%
   \binrel@{#1}\binrel@@{#2}}
\def\binrel@#1{\setboxz@h{\thinmuskip0mu
  \medmuskip\m@ne mu\thickmuskip\@ne mu$#1\m@th$}%
 \setbox@ne\hbox{\thinmuskip0mu\medmuskip\m@ne mu\thickmuskip
  \@ne mu${}#1{}\m@th$}%
 \setbox\tw@\hbox{\hskip\wd@ne\hskip-\wdz@}}
\def\binrel@@#1{\ifdim\wd2<\z@\mathbin{#1}\else\ifdim\wd\tw@>\z@
 \mathrel{#1}\else{#1}\fi\fi}
\def\mathch{\protect\p@mathch}
\def\p@mathch#1#2{%
  \begingroup
  \let\@nomath\@gobble \mathversion{#1}%
  \math@atom{#2}{%
    \mathchoice%
    {\hbox{$\m@th\displaystyle#2$}}%
    {\hbox{$\m@th\textstyle#2$}}%
    {\hbox{$\m@th\scriptstyle#2$}}%
    {\hbox{$\m@th\scriptscriptstyle#2$}}}%
  \endgroup}
\def\bmath{\protect\p@boldm}
\def\p@boldm#1{\mathch{bold}{#1}}
\makeatother

\defcitealias{Kelly0705.2774}{K07}
\newcommand*{\koseven}{\citetalias{Kelly0705.2774}}

\begin{document}

\title{A Gibbs Sampler for Multivariate Linear Regression}

\author[A. B. Mantz]{Adam B. Mantz$^{1,2,3,4}$\thanks{Corresponding author e-mail: \href{mailto:amantz@slac.stanford.edu}{\tt amantz@slac.stanford.edu}} {} \\
  $^1$Department of Astronomy and Astrophysics, University of Chicago, 5640 South Ellis Avenue, Chicago, IL 60637, USA\\
  $^2$Kavli Institute for Cosmological Physics, University of Chicago, 5640 South Ellis Avenue, Chicago, IL 60637, USA\\
  $^3$Kavli Institute for Particle Astrophysics and Cosmology, Stanford University, 452 Lomita Mall, Stanford, CA 94305, USA\\
  $^4$Department of Physics, Stanford University, 382 Via Pueblo Mall, Stanford, CA 94305, USA\\
}
\date{Submitted 18 August 2015. Accepted 24 December 2015.}

\maketitle

\begin{abstract}
  \citet[][hereafter \koseven{}]{Kelly0705.2774} described an efficient algorithm, using Gibbs sampling, for performing linear regression in the fairly general case where non-zero measurement errors exist for both the covariates and response variables, where these measurements may be correlated (for the same data point), where the response variable is affected by intrinsic scatter in addition to measurement error, and where the prior distribution of covariates is modeled by a flexible mixture of Gaussians rather than assumed to be uniform. Here I extend the \koseven{} algorithm in two ways. First, the procedure is generalized to the case of multiple response variables. Second, I describe how to model the prior distribution of covariates using a Dirichlet process, which can be thought of as a Gaussian mixture where the number of mixture components is learned from the data. I present an example of multivariate regression using the extended algorithm, namely fitting scaling relations of the gas mass, temperature, and luminosity of dynamically relaxed galaxy clusters as a function of their mass and redshift. An implementation of the Gibbs sampler in the {\sc r} language, called {\sc lrgs}, is provided.
\end{abstract}


\section{Introduction}

Linear regression is perhaps the most widely used example of parameter fitting throughout the sciences. Yet, the traditional ordinary least-squares (or weighted least-squares) approach to regression neglects some features that are practically ubiquitous in astrophysical data, namely the existence of measurement errors, often correlated with one another, on \emph{all} quantities of interest, and the presence of residual, intrinsic scatter (i.e.\ physical scatter, not the result of measurement errors) about the best fit. \koseven{} takes on this problem (see that work for a more extensive overview of the prior literature) by devising an efficient algorithm for simultaneously constraining the parameters of a linear model and the intrinsic scatter in the presence of such heteroscedastic and correlated measurement errors. In addition, the \koseven{} approach corrects a bias that exists when the underlying distribution of covariates in a regression is assumed to be uniform, by modeling this distribution as a flexible mixture of Gaussian (normal) distributions and marginalizing over it.

The \koseven{} model is considerably more complex, in terms of the number of free parameters, than traditional regression. Nevertheless, it can be efficiently constrained using a fully conjugate Gibbs sampler, as described in that work. Briefly, the approach takes advantage of the fact that, for a suitable model, the fully conditional posterior of certain parameters (or blocks of parameters)\footnote{i.e.\ the posterior distribution for certain parameters conditional on the (fixed) values of all other parameters.}  may be expressible as a known distribution which can be sampled from directly using standard numerical techniques. If all model parameters can be sampled this way, then a Gibbs sampler, which simply cycles through the list of parameters, updating or block-updating them in turn, can move efficiently through the parameter space. By repeatedly Gibbs sampling, a Markov chain that converges to the joint posterior distribution of all model parameters is generated (see, e.g., \citealt{Gelman2004BayesianDataAnalysis} for theoretical background). The individual pieces (e.g., the model distributions of measurement error, intrinsic scatter, and the covariate prior distribution) of the \koseven{} model are conjugate, making it suitable for this type of efficient Gibbs sampling. This is a key advantage in terms of making the resulting algorithm widely accessible to the community, since conjugate Gibbs samplers, unlike more general and powerful Markov Chain Monte Carlo samplers, require no a priori tuning by the user.

While \koseven{} argue against the assumption of a uniform prior for covariates, it should be noted that the alternative of a Gaussian mixture model (or the Dirichlet process generalization introduced below) is not necessarily applicable in every situation either. When a well motivated physical model of the distribution of covariates exists, it may well be preferable to use it, even at the expense of computational efficiency. In the general case, we can hope that a flexible parametrization like the Gaussian mixture is adequate, although it is always worth checking a posteriori that the model distribution of covariates provides a good description of the data. \koseven{} and \citet{Sereno1502.05413} discuss real applications in which a Gaussian distribution of covariates turns out to be adequate, despite the underlying physics being non-Gaussian.

This work describes two useful generalizations to the \koseven{} algorithm. First, the number of response variables is allowed to be greater than one. Second, the prior distribution of covariates may be modeled using a Dirichlet process rather than as a mixture of Gaussians with a fixed number of components. A Dirichlet process describes a probability distribution over the space of probability distributions, and (in contrast to the many parameters required to specify a large mixing model) is described only by a concentration parameter and a base distribution. For the choice of a Gaussian base distribution, used here, the Dirichlet process can be thought of as a Gaussian mixture in which the number of mixture components is learned from the data and marginalized over as the fit progresses (see more discussion, in a different astrophysical context, by \citealt{Schneider1411.2608}). This makes it a very general and powerful alternative to the standard fixed-size Gaussian mixture, as well as one that requires even less tuning by the user, since the number of mixture components need not be specified. Crucially, both of these generalizations preserve the conjugacy of the model, so that posterior samples can still be easily obtained by Gibbs sampling.

Of course, \koseven{} (or this paper) does not provide the only implementation of conjugate Gibbs sampling, nor is that approach the only one possible for linear regression in the Bayesian context. Indeed, there exist more general statistical packages capable of identifying conjugate sampling strategies (where possible) based on an abstract model definition (e.g., {\sc bugs},\footnote{\url{http://openbugs.net/w/FrontPage}} {\sc jags},\footnote{\url{http://mcmc-jags.sourceforge.net/}} and {\sc stan}\footnote{\url{http://mc-stan.org/}}). The use of more general Markov chain sampling techniques naturally allow for more general (non-conjugate) models and/or parametrizations (e.g., \citealt{Maughan1212.0858, Robotham1508.02145}). Nevertheless, there is something appealing in the relative simplicity of implementation and use of the conjugate Gibbs approach, particularly as it applies so readily to the commonly used linear model with Gaussian scatter.

\secref~\ref{sec:model} describes the model employed in this work in more detail, and introduces notation. \secref~\ref{sec:sampler} outlines the changes to the \koseven{} sampling algorithm needed to accomodate the generalizations above. Since this work is intended to extend that of \koseven{}, I confine this discussion only to steps which differ from the that algorithm, and do not review the Gibbs sampling procedure in its entirety. However, the level of detail is intentionally high; between this document and \koseven{}, it should be  straightforward for the interested reader to create his or her own implementation of the entire algorithm. \secref~\ref{sec:examples} provides some example analyses, including one with real astrophysical data, and discusses some practical aspects of the approach.

The complete algorithm described here (with both Gaussian mixture and Dirichlet process models) has been implemented in the {\sc r} language.\footnote{\url{http://www.r-project.org}} The package is named Linear Regression by Gibbs Sampling ({\sc lrgs}), the better to sow confusion among extragalactic astronomers. The code can be obtained from GitHub\footnote{\url{https://github.com/abmantz/lrgs}} or the Comprehensive R Archive Network.\footnote{\url{http://cran.r-project.org}}

\section{Model and Notation} \label{sec:model}

Here I review the model described by \koseven{}, introducing the generalization to multiple response variables (\secref~\ref{sec:mvmodel}) and the use of the Dirichlet process to describe the prior distribution of the covariates (\secref~\ref{sec:dproc}). The notation used here is summarized in \tabref~\ref{tab:notation}; it differs slightly from that of \koseven{}, as noted. In this document, $A\sim B$ denotes a stochastic relationship in which a random variable $A$ is drawn from the probability distribution $B$, and boldface distinguishes vector- or matrix-valued variables.

\begin{table*}
  \centering
  \caption[]{Summary of notation used in this work. Where this departs from the notation used by \koseven{}, the \koseven{} equivalent is noted in the last column.}
  \vspace{1ex}
  \begin{tabular}{lclc}
    \hline
    & Symbol & Meaning & \koseven{} \\
    \hline
    General
    &$\normal_\nu(\bmath{\mu},\bmath{\Sigma})$ & $\nu$-dimensional normal distribution (mean $\bmath{\mu}$, covariance $\bmath{\Sigma}$) \\
    notation
    &$\wishart(\bmath{V},\nu)$ & Wishart distribution (scale matrix $\bmath{V}$, $\nu$ degrees of freedom) \\
    &$A_{ij}$ & single element of matrix $\bmath{A}$ \\
    &$\bmath{A_{j\cdot}},\bmath{A_{\cdot j}}$ & $j$th row or column of $\bmath{A}$ \\
    &$\bmath{A_{\bar{j}\cdot}},\bmath{A_{\cdot \bar{j}}}$ & $\bmath{A}$ with the $j$th row or column removed & $\bmath{A_{-j\cdot}},\bmath{A_{\cdot-j}}$ \\
    &$\one_n$ & $n\times n$ identity matrix \\
    \hline
    Common
    &$n$ & number of data points \\
    parameters
    &$p$ & number of covariates \\
    &$m$ & number of responses & 1 \\
    &$K$ & number of Gaussian mixture components or clusters \\
    &$\bmath{x_i},\bmath{y_i}$ & measured covariates and responses for data point $i$ \\
    &$\bmath{M_i}$ & measurement covariance matrix for data point $i$ & $\bmath{\Sigma_i}$ \\
    &$\bmath{\xi_i},\bmath{\eta_i}$ & true covariates and responses for data point $i$ \\
    &$\bmath{\alpha}$ & intercepts of the linear model \\
    &$\bmath{\beta}$ & slopes of the linear model \\
    &$\bmath{\Sigma}$ & intrinsic covariance about the linear model & $\sigma^2$ \\
    &$\bmath{G}$ & mixture component/cluster identification for each data point \\
    \hline
    Gaussian
    &$\bmath{\pi}$ & weight of the each mixture component \\
    mixture
    &$\bmath{\mu_k}$ & mean of the $k$th component \\
    &$\bmath{T_k}$ & covariance of the $k$th component \\
    &$\bmath{\mu_0}$ & mean of the prior distribution of each $\bmath{\mu_k}$ \\
    &$\bmath{U}$ & covariance of the prior distribution of each $\bmath{\mu_k}$ \\
    &$\bmath{W}$ & scale matrix of the prior distribution of each $\bmath{T_k}$ \\
    \hline
    Dirichlet
    &$\bmath{\mu}$ & mean of the normal base distribution \\
    process
    &$\bmath{T}$ & covariance of the normal base distribution \\
    &$\kappa$ & concentration parameter of the process \\
    &$a$ & shape parameter of the prior of $\kappa$ \\
    &$b$ & rate parameter of the prior of $\kappa$ \\
    \hline
  \end{tabular}
  \label{tab:notation}
\end{table*}

\subsection{Gaussian mixture model} \label{sec:mvmodel}

We are interested in $p+m$ properties of some class of object, where $p$ of these (covariates) are supposed to be physically responsible for determining the other $m$ (response variables). Measurements of these $p+m$ quantities have been gathered for $n$ objects. The true values of the covariates for the $i$th data point are denoted $\bmath{\xi_i}$, and the corresponding true responses are denoted $\bmath{\eta_i}$; these are nuisance parameters that will be marginalized over. The measured values of the corresponding quantities are denoted $\bmath{x_i}$ and $\bmath{y_i}$, and are assumed to be related to the true values by a $(p+m)$-dimensional normal measurement error distribution, which may be different for each data point. Writing the $\nu$-dimensional normal distribution with mean $\bmath{\mu}$ and covariance $\bmath{V}$ as $\normal_\nu(\bmath{\mu},\bmath{V})$, this is (for the $i$th data point)\footnote{Note that in \koseven{} $\bmath{y}$ preceded $\bmath{x}$ as they correspond to rows and columns of $\bmath{M}$. The reverse convention is followed here.}
\begin{equation} \label{eq:datamodel}
  \left(\begin{array}{c}\bmath{x_i}\\\bmath{y_i}\end{array}\right) \sim \normal_{p+m}\left[\left(\begin{array}{c}\bmath{\xi_i}\\\bmath{\eta_i}\end{array}\right),\, \bmath{M_i}\right].
\end{equation}

The $p$-dimensional distribution of covariates that these objects originally come from is not necessarily uniform. It is therefore modeled in a flexible way, as a mixture of $K$ $p$-dimensional normal distributions,
\begin{equation} \label{eq:mixture}
  \bmath{\xi_i} \sim \sum_{k=1}^{K} \pi_k \, \normal_p\left( \bmath{\mu_k},\, \bmath{T_k} \right),
\end{equation}
with $\sum_k \pi_k=1$. The summation notation in \eqnref~\ref{eq:mixture} is meant to convey that $\bmath{\xi_i}$ is drawn from the $k$th normal distribution (which has mean $\bmath{\mu_k}$ and covariance $\bmath{T_k}$) with probability $\pi_k$. As in \koseven{}, this is implemented by means of a set of latent indicator variables, $\bmath{G}$, with $G_i$ indicating which of the $K$ mixture components $\bmath{\xi_i}$ is drawn from.\footnote{In the notation used here, $G_i$ is simply a label $1,2,\ldots,K$, whereas \koseven{} describe each $G_i$ as a vector with all but one element zero. This distinction makes no practical difference.} Formally, each $\bmath{G}$ follows the multinomial distribution defined by the proportions $\bmath{\pi}$.

The parameters $\bmath{G}$, $\bmath{\pi}$, $\{\bmath{\mu_k}\}$ and $\{\bmath{T_k}\}$ can be learned from the data, but it is helpful to impose some structure on them. Therefore, we adopt a hierarchical model whereby the vectors $\{\bmath{\mu_k}\}$ themselves follow a normal distribution,
\begin{equation} \label{eq:mu}
  \bmath{\mu_k} \sim \normal_p\left( \bmath{\mu_0},\, \bmath{U} \right),
\end{equation}
and both $\bmath{U}$ and the covariances $\{\bmath{T_k}\}$ follow inverse-Wishart distributions,
\begin{eqnarray} \label{eq:UTau}
  \bmath{U} &\sim& \wishart^{-1}\left(\bmath{W},K+p\right), \\
  \bmath{T_k} &\sim& \wishart^{-1}\left(\bmath{W},\, K+p\right).\nonumber
\end{eqnarray}
Here $\wishart^{-1}(\bmath{V}, \nu)$ denotes the inverse-Wishart distribution with scale matrix $\bmath{V}$ and $\nu$ degrees of freedom.\footnote{While the inverse-Wishart distribution is conjugate in this context, and therefore computationally convenient, it has the generic disadvantage of imposing a particular structure on the marginal prior distributions of the variances and correlation coefficients that make up the resulting covariance matrix. In particular, large absolute values of the correlation coefficients preferentially correspond to large variances. This is sometimes undesirable, and from a practical standpoint it can occasionally result in the generation of computationally singular matrices. An alternative approach is to decompose a given covariance matrix, $\bmath{\Lambda}$, as $\bmath{SRS}$, where $\bmath{S}$ is diagonal and $\bmath{R}$ is a correlation matrix. This allows independent priors to be adopted for individual variances and correlation coefficients, which is usually more intuitive than using inverse-Wishart distribution, but at the expense that the resulting model is no longer conjugate. More extensive discussion of these options can be found in \citet{Barnard24306780}, \citet{Gelman2004BayesianDataAnalysis}, and \cite{OMalley27640192}.} I follow \koseven{} in taking uniform priors on the hyperparameters $\bmath{\mu_0}$ and $\bmath{W}$.\footnote{The conjugate prior for $\bmath{\mu_0}$ is normal, $\normal_p(\bmath{u},\bmath{V})$, which is uniform in the limit $\bmath{V^{-1}}=\bmath{0}$. For the $\bmath{W}$, the conjugate prior is inverse-Wishart, $\wishart^{-1}(\bmath{\Psi}, \nu)$, and the equivalent of the uniform distribution is realized by taking $\bmath{\Psi}=\bmath{0}$ and $\nu=-(1+d)$, where $d$ is the size of $\bmath{\Psi}$ (in this case, $d=p$).} Note that this hierarchical model, and the Gaussian mixture itself, is fairly flexible but not fully general. While these are typically reasonable assumptions when we have little prior information about the distribution of covariates, they may not be appropriate for all situations. The particular structure in Equations~\ref{eq:mu}--\ref{eq:UTau} tends to promote compactness in the covariate distribution; that is, if multiple, well separated clusters of covariates exist, the onus is on the data to show that they are required.

The relationship by which the $p$ covariates determine the $m$ responses is assumed to be linear, with a normal intrinsic scatter,
\begin{equation} \label{eq:linearmodel}
  \bmath{\eta_i} \sim \normal_m\left(\bmath{\alpha} + \bmath{\beta} \bmath{\xi_i},\, \bmath{\Sigma}\right).
\end{equation}
Note that the linearity of the mean is crucial to maintaining the conjugacy of the model. Here $\bmath{\alpha}$ is the $m\times1$ vector of intercepts, $\bmath{\beta}$ is an $m\times p$ matrix of slopes linking each response variable with each of the covariates, and $\bmath{\Sigma}$ is the $m\times m$ intrinsic covariance matrix (assumed to be constant with respect to $\bmath{\xi}$). This can be written compactly for the entire data set in matrix form, with the definitions
\begin{eqnarray}
  \bmath{X_{i\cdot}} &=& (1,\, \transpose{\bmath{\xi_i}}), \\
  \bmath{Y_{i\cdot}} &=& \transpose{\bmath{\eta_i}}, \nonumber\\
  \bmath{B} &=& \transpose{( \bmath{\alpha},\, \bmath{\beta} )}. \nonumber
\end{eqnarray}
where $\bmath{Y}$ is $n\times m$, $\bmath{X}$ is $n\times(p+1)$, and $\bmath{B}$ is $(p+1)\times m$. The notation $\bmath{A_{i\cdot}}$ refers to the $i$th row of $A$; likewise $\bmath{A_{\cdot j}}$ would refer to the $j$th column. The statement of the linear model then takes the familiar form
\begin{eqnarray}\label{eq:alldata} \label{eq:lm}
  \bmath{Y} &=& \bmath{X B} + \bmath{E}, \\
  \bmath{E_{i\cdot}} &\sim& \normal_m\left(\bmath{0},\, \bmath{\Sigma}\right). \nonumber
\end{eqnarray}
As noted in Sections~\ref{sec:coefficients} and \ref{sec:intrinsic}, the conjugate prior distributions for $\bmath{B}$ and $\bmath{\Sigma}$ are, respectively, normal and inverse-Wishart.

\subsection{Dirichlet process model} \label{sec:dproc}

The Gaussian mixture prior on the distribution of covariates is flexible, but requires us to either chose a number of mixture components outright or carefully check that the sensitivity of results to the number of components. Alternatively, we can constrain the distribution of covariates using a Dirichlet process, which describes a probability distribution over probability distributions. A Dirichlet process is defined by a concentration parameter, $\kappa$, and a base distribution, $P_0$. By choosing $P_0$ to be $p$-dimensional normal, the conjugacy relations that made the Gaussian mixture efficient to Gibbs sample will also hold for the Dirichlet process. Used in this way, the Dirichlet process can be thought of as a Gaussian mixture in which the number of components is marginalized over (\citealt{Neal1390653} and references therein). The analog of \eqnref~\ref{eq:mixture} is generally written
\begin{eqnarray}
  \bmath{\xi_i} &\sim& P, \\
  P &\sim& \mathrm{DP}(P_0, \kappa), \nonumber\\
  P_0 &=& \normal_p(\bmath{\mu}, \bmath{T}). \nonumber
\end{eqnarray}
Here $\bmath{\mu}$ and $\bmath{T}$ are the hyperparameters of the base distribution, for which I assume uniform priors. Note that there is only one $\bmath{\mu}$ and one $\bmath{T}$, unlike in the Gaussian mixture model, and there is no analog of $\bmath{\mu_0}$, $\bmath{U}$ or $\bmath{W}$. The remainder of the model, namely \eqnref{}s~\ref{eq:datamodel} and \ref{eq:linearmodel}--\ref{eq:lm}, is the same as above.

In practice, for a given realization of the model parameters, the algorithm for realizing the Dirichlet process divides the data set into a finite number of clusters, with points in each cluster having identical values of $\bmath{\xi}$.\footnote{A pitfall of this approach occurs when few of the measured covariates are consistent with any others within their measurement errors. In that case, the number of clusters is necessarily similar in size to the number of data points, which is not generally the desired result.} The vector of labels $\bmath{G}$ will identify which cluster each data point belongs to, similarly to its use in \secref~\ref{sec:mvmodel}. A vector of cluster proportions, $\bmath{\pi}$, could also be defined analogously. However, in practice, the procedure for Gibbs sampling the Dirichlet process model implicitly marginalizes over it, and so $\bmath{\pi}$ never explicitly appears in the calculations (\secref~\ref{sec:dpgibbs}).

The concentration parameter of the Dirichlet process is related to the number of clusters in the data set, and can also be marginalized over. The conjugate prior for $\kappa$ is the Gamma distribution,
\begin{equation}
  \kappa \sim \mathrm{Gamma}(a, b),
\end{equation}
where $a$ and $b$ are respectively the shape and rate parameters of the prior. If the approximate number of clusters in the data set is known, these parameters can be chosen accordingly; otherwise, they can be chosen to be minimally informative (see discussion by \citealt{Dorazio20093384} and \citealt{Murugiah2012.02.013}, and \secref~\ref{sec:toy}).

\section{The Gibbs Sampler} \label{sec:sampler}

Both of the models described above can be efficiently Gibbs sampled because they are fully conjugate. Recall that, in this situation, the sampling algorithm can be entirely specified as the set of conditional distributions used to sequentially update each parameter or block of parameters. \secref~\ref{sec:gmgibbs} summarizes the changes to the \koseven{} procedure needed to sample the Gaussian mixture model when there are multiple response variables, and \secref~\ref{sec:dpgibbs} describes the procedure for sampling the Dirichlet process model.

In either case, an initial guess is needed for most of the free parameters, but this need not be very sophisticated. For example, it is generally acceptable to begin with the values of $\{\bmath{\xi_i}\}$ and $\{\bmath{\eta_i}\}$ respectively initialized to the measured values $\{\bmath{x_i}\}$ and $\{\bmath{y_i}\}$, the intercepts $\bmath{\alpha}$ set to the average value of each column of $\bmath{Y}$, and the slopes $\bmath{\beta}$ set to zero.\footnote{Specifically, this simpleminded guess for the intercepts and slopes works reasonably well when the covariates have been approximately centered. More generally, estimates from an ordinary least-squares regression should provide a good starting point.} Of course, more intelligent guesses will decrease the ``burn-in'' time of the resulting Markov chain, but generally this is relatively short.

\subsection{Sampling the Gaussian mixture model} \label{sec:gmgibbs}

The procedure for updating the parameters governing the distribution of covariates ($\bmath{G}$, $\bmath{\pi}$, $\{\bmath{\mu_k}\}$, $\{\bmath{T_k}\}$, $\bmath{\mu_0}$, $\bmath{U}$ and $\bmath{W}$, for which I adopt the same priors as \koseven{}) is not affected by the generalization to multiple responses, and the reader is referred to \koseven{} for the details of those updates. Here, I review the procedure for Gibbs sampling the true values of the covariates and responses for each data point, $\{\bmath{\xi_i}\}$ and $\{\bmath{\eta_i}\}$, the regression coefficients, $\bmath{\alpha}$ and $\bmath{\beta}$, and the intrinsic covariance matrix, $\bmath{\Sigma}$.

\subsubsection{Updating the covariates}\label{sec:covariates}

See Equations~59--65 of \koseven{} for the corresponding discussion in that work. The fully conditional posterior of the $j$th covariate ($j=1,2,\ldots,p$) for data point $i$ is
\begin{eqnarray}
  (\bmath{\xi_i})_j | \ldots \sim \normal_1\left(\hat{\xi}_{(ij)},\sigma _{(ij)}^2\right),
\end{eqnarray}
where 
\begin{eqnarray}
  \sigma _{(ij)}^2 &=& \left[ (\bmath{M_i}^{-1})_{jj} + (\bmath{T_{G_i}})^{-1}_{jj} + \transpose{\bmath{\beta_{\cdot j}}}\bmath{\Sigma}^{-1}\bmath{\beta_{\cdot j}} \right]^{-1}, \\
  \hat{\xi}_{(ij)} &=& \sigma _{(ij)}^2\left[ \left(\bmath{M_i}^{-1}\bmath{z_i^{*}}\right)_j + \left(\bmath{T_{G_i}}^{-1}\bmath{\mu_i^{*}}\right)_j \right. \nonumber\\
    && \left. + \transpose{\bmath{\beta_{\cdot j}}}\bmath{\Sigma}^{-1}\left(\bmath{\eta_i}-\bmath{\alpha}-\bmath{\beta_{\cdot\bar{j}}}(\bmath{\xi_i})_{\bar{j}}\right) \right]. \nonumber
\end{eqnarray}
Here $\bar{j}$ indicates removal of the $j$th entry or column, and $\bmath{z_i^{*}}$ and $\bmath{\mu_i^{*}}$ are defined as in \koseven{},
\begin{eqnarray}
  (\bmath{z_i^*})_\ell &=& \left\{\begin{array}{r} (\bmath{x_i})_\ell, \quad \ell = j \\ (\bmath{x_i},\bmath{y_i})_\ell - (\bmath{\xi_i},\bmath{\eta_i})_\ell, \quad \ell \neq j\end{array}\right., \\
  (\bmath{\mu_i^*})_\ell &=& \left\{ \begin{array}{r} (\bmath{\mu_{G_i}})_\ell, \quad \ell=j\\ (\bmath{\mu_{G_i}})_\ell - (\bmath{\xi_i})_\ell, \quad \ell \neq j\end{array}\right..\nonumber
\end{eqnarray}

\subsubsection{Updating the responses}\label{sec:responses}

The response variables, $\{\bmath{\eta_i}\}$, can be updated by modifying Equations~69--72 of \koseven{} as follows (for each $j=1,2,\ldots,m$):
\begin{eqnarray}
  (\bmath{\eta_i})_j | \ldots &\sim& \normal_1\left(\hat{\eta}_{(ij)},s _{(ij)}^2\right), \\
  s_{(ij)}^2 &=& \left[ \left(\bmath{M_i}^{-1}\right)_{(p+j)(p+j)} + \Sigma^{-1}_{jj} \right]^{-1}, \nonumber\\
  \hat{\eta}_{(ij)} &=& s _{(ij)}^2\left[ \left(\bmath{M_i}^{-1}\bmath{\zeta_i^{*}}\right)_{p+j} + \left(\bmath{\Sigma}^{-1}\bmath{q_i^{*}}\right)_j \right]. \nonumber
\end{eqnarray}
Here $\bmath{\zeta_i^{*}}$ is defined analogously $\bmath{z_i^{*}}$ in \secref~\ref{sec:covariates},
\begin{eqnarray}
  (\bmath{\zeta_i^*})_\ell &=& \left\{\begin{array}{r} (\bmath{y_i})_\ell, \quad \ell = p+j \\ (\bmath{x_i},\bmath{y_i})_\ell - (\bmath{\xi_i},\bmath{\eta_i})_\ell, \quad \ell \neq p+j\end{array}\right., \nonumber\\
\end{eqnarray}
and
\begin{eqnarray}
  \left(\bmath{q_i^{*}}\right)_\ell = \Bigg\{ \begin{array}{r}\alpha_\ell+\bmath{\beta_{\ell\cdot}}\bmath{\xi_i}, \quad \ell=j\\\alpha_\ell+\bmath{\beta_{\ell\cdot}}\bmath{\xi_i}-(\bmath{\eta_i})_\ell, \quad \ell\neq j\end{array}.
\end{eqnarray}

\subsubsection{Updating the coefficients}\label{sec:coefficients}

The coefficients, $\bmath{\alpha}$ and $\bmath{\beta}$, may be updated by recasting Equation~\ref{eq:alldata} in the form of a univariate regression,
\begin{eqnarray}
  \bmath{\widetilde{Y}} = \bmath{\widetilde{X}\widetilde{B}} + \bmath{\widetilde{E}},
\end{eqnarray}
where $\bmath{\widetilde{Y}}$ and $\bmath{\widetilde{E}}$ are $nm\times1$, $\bmath{\widetilde{X}}$ is $nm\times (p+1)m$ and $\bmath{\widetilde{B}}$ is $(p+1)m\times1$. I use the following (non-unique) definitions:
\begin{eqnarray}
  \bmath{\widetilde{Y}} &=& \left(\begin{array}{c}\bmath{Y_{\cdot1}}\\\vdots\\\bmath{Y_{\cdot m}}\end{array}\right), \\
  \bmath{\widetilde{B}} &=& \left(\begin{array}{c}\bmath{B_{\cdot1}}\\\vdots\\\bmath{B_{\cdot m}}\end{array}\right), \nonumber\\
  \bmath{\widetilde{X}} &=& \left(\begin{array}{ccc}\bmath{X}&\bmath{0}&\cdots\\\bmath{0}&\bmath{X}&\cdots\\\vdots&\vdots&\ddots\end{array}\right), \nonumber
\end{eqnarray}
with the $nm\times nm$ scatter covariance being
\begin{eqnarray}
  \bmath{\widetilde{\Sigma}} &=& \left(\begin{array}{cccc}\one_n\Sigma_{11}&\one_n\Sigma_{12}&\cdots&\one_n\Sigma_{1m}\\\one_n\Sigma_{21}&\one_n\Sigma_{22}&\cdots&\one_n\Sigma_{2m}\\\vdots&\vdots&\ddots&\vdots\\\one_n\Sigma_{m1}&\one_n\Sigma_{m2}&\cdots&\one_n\Sigma_{mm}\end{array}\right),
\end{eqnarray}
where $\one_n$ denotes the $n\times n$ identity. The fully conditional posterior for $\bmath{\widetilde{B}}$ is simply the normal distribution following from ordinary least-squares regression, 
\begin{eqnarray} \label{eq:coeffsample}
  \bmath{\widetilde{B}} | \ldots \sim \normal_{(p+1)m}(\bmath{\widetilde{\mathcal{B}}},\bmath{\widetilde{\mathcal{S}}}).
\end{eqnarray}
 The mean is
\begin{eqnarray}
  \bmath{\widetilde{\mathcal{B}}} = \left(\transpose{\bmath{\widetilde{X}}}\bmath{\widetilde{X}}\right)^{-1}\transpose{\bmath{\widetilde{X}}}\bmath{\widetilde{Y}},
\end{eqnarray}
whose calculation can be broken down into $(p+1)\times(p+1)$ chunks due to the structure of $\bmath{\widetilde{X}}$, and the covariance is
\begin{eqnarray}
  \bmath{\widetilde{\mathcal{S}}} &=& \left(\begin{array}{cccc}\bmath{\Xi} \, \Sigma_{11}&\bmath{\Xi} \, \Sigma_{12}&\cdots&\bmath{\Xi} \, \Sigma_{1m}\\\bmath{\Xi} \, \Sigma_{21}&\bmath{\Xi} \, \Sigma_{22}&\cdots&\bmath{\Xi} \, \Sigma_{2m}\\\vdots&\vdots&\ddots&\vdots\\\bmath{\Xi} \, \Sigma_{m1}&\bmath{\Xi} \, \Sigma_{m2}&\cdots&\bmath{\Xi} \, \Sigma_{mm}\end{array}\right), \\
  \bmath{\Xi} &=& \left(\transpose{\bmath{\widetilde{X}}}\bmath{\widetilde{X}}\right)^{-1} \nonumber
\end{eqnarray}
i.e. $\mathrm{Cov}\left(B_{ki},B_{\ell j}\right) = \Sigma_{ij}\left(\transpose{\bmath{\widetilde{X}}}\bmath{\widetilde{X}}\right)^{-1}_{k\ell}.$

Note that it is straightforward to sample from the product of \eqnref~\ref{eq:coeffsample} and a normal prior for $\bmath{\widetilde{B}}$; this option is implemented in {\sc lrgs}, although the default is a uniform prior.

\subsubsection{Updating the intrinsic covariance}\label{sec:intrinsic}

With $\bmath{E} = \bmath{Y} - \bmath{XB}$ (Equation~\ref{eq:alldata}), the conditional posterior for the intrinsic scatter is\footnote{This expression assumes a Jeffreys (i.e., minimally informative) prior on $\Sigma$. More generally, one could use a prior $\bmath{\Sigma} \sim \wishart^{-1}(\bmath{\Psi}, \nu_0)$, in which case the conditional posterior becomes $\bmath{\Sigma} | \ldots \sim \wishart^{-1}\left(\transpose{\bmath{E}}\bmath{E}+\bmath{\Psi},\, n+\nu_0\right)$. The Jeffreys prior corresponds to $\bmath{\Psi}=\bmath{0}$ and $\nu_0=-1$, while $\bmath{\Psi}=\bmath{0}$ and $\nu_0=-(1+m)$ corresponds to a prior that is uniform in $\left|\bmath{\Sigma}\right|$ (see, e.g., \citealt{Gelman2004BayesianDataAnalysis}). The \koseven{} algorithm makes the latter assumption. The default in {\sc lrgs} is the Jeffreys prior, but any inverse-Wishart prior can optionally be specified.}
\begin{eqnarray} \label{eq:scattersample}
  \bmath{\Sigma} | \ldots \sim \wishart^{-1}\left(\transpose{\bmath{E}}\bmath{E},\, n-1\right).
\end{eqnarray}

In practice, a sample can be generated by setting $\bmath{\Sigma}$ equal to $\left(\transpose{\bmath{A}}\bmath{A}\right)^{-1}$, where $\bmath{A}$ is $(n-1) \times m$ and each row of $\bmath{A}$ is generated as $\bmath{A_{i\cdot}}\sim\normal_m\left[\bmath{0},\left(\transpose{\bmath{E}}\bmath{E}\right)^{-1}\right]$.

\subsection{Sampling the Dirichlet process model} \label{sec:dpgibbs}

If a Dirichlet process rather than a Gaussian mixture is used to describe the prior distribution of covariates, the procedure to update the $\{\bmath{\xi_i}\}$ differs from that given above. This step implicitly updates $\bmath{G}$, which now identifies membership in one of a variable number of clusters (data points with identical values of $\bmath{\xi}$). In addition, Gibbs updates to the hyperparameters of the Dirichlet process and its base distribution, $\kappa$, $\bmath{\mu}$ and $\bmath{T}$, are possible. These are described below. Note that the updates to $\bmath{G}$, $\bmath{\pi}$, $\{\bmath{\mu_k}\}$, $\{\bmath{T_k}\}$, $\bmath{\mu_0}$, $\bmath{U}$ and $\bmath{W}$ given in \koseven{} are no longer applicable (of these, only $\bmath{G}$ and $\bmath{\pi}$ exist in the model).

\subsubsection{Updating the covariates} \label{sec:dpcovariates}

Let $K$ be the number of clusters (i.e.\ distinct labels in $\bmath{G}$) at a given time. I follow the second algorithm given by \citet{Neal1390653}, which first updates the cluster membership for each data point, and then draws new values of $\bmath{\xi}$ for each cluster.

For each data point $i$, update $G_i$ as follows. Let
\begin{equation}
  q_k^{(i)} = n_k^{(i)} \, \normal_p\left(\bmath{\xi'_k} | \bmath{\hat{\xi}_1^{(i)}}, \bmath{\hat{T}_1^{(i)}}\right), \quad k=1,2,\ldots,K,
\end{equation}
where $n_k$ is the number of data points belonging to the $k$th cluster \emph{not counting} the $i$th data point, $\bmath{\xi'_k}$ is the vector of covariates shared by the $k$th cluster, and $\normal_\nu(\bmath{x}|\bmath{\mu},\bmath{V})$ denotes the normal density, i.e.\ the density of $\normal_\nu(\bmath{\mu},\bmath{V})$ evaluated at $\bmath{x}$. Here
\begin{eqnarray}
  \bmath{\hat{T}_1^{(i)}} &=& \left[ (\bmath{M_i}^{-1})_{xx} + \transpose{\bmath{\beta}}\bmath{\Sigma}^{-1}\bmath{\beta} \right]^{-1}, \\
  \bmath{\hat{\xi}_1^{(i)}} &=&  \bmath{\hat{T}_1^{(i)}} \left[ \left(\bmath{M_i}^{-1}\bmath{z_i}\right)_x + \transpose{\bmath{\beta}}\bmath{\Sigma}^{-1}\left(\bmath{\eta_i}-\bmath{\alpha}\right) \right], \nonumber
\end{eqnarray}
where $\bmath{z_i} = \left( \bmath{x_i} , \bmath{y_i}-\bmath{\eta_i} \right)$, and the subscript $x$ indicates the range of subscripts associated with the covariates, $1,2,\ldots,p$ (so that, e.g.,  $ [\bmath{M_i}^{-1}]_{xx}$ is the upper-left $p\times p$ block of $\bmath{M_i^{-1}}$). Furthermore, let
\begin{equation}
  r^{(i)} = \kappa \, \normal_p\left( \bmath{\mu} \left| \bmath{\hat{\xi}_1^{(i)}},\, \bmath{\hat{T}_1^{(i)}} + \bmath{T} \right.\right).
\end{equation}
Each element of $\bmath{q^{(i)}}$ is the conditional probability associated with the covariates of the $k$th cluster given the measurement and response variables associated with the $i$th data point, whereas $r^{(i)}$ is related to the probability of the $i$th data point being drawn instead from the base distribution of the Dirichlet process. A new label, $G_i$, is drawn from the multinomial distribution as
\begin{equation}
  G_i|\ldots \sim \mathrm{Multinom}\left[(\bmath{q^{(i)}},r^{(i)})\right],
\end{equation}
after normalizing the probability vector $(\bmath{q^{(i)}},r^{(i)})$. A selection $G_i=K+1$ indicates the creation of a new cluster, and in that case a new $\bmath{\xi_i}$ is immediately drawn from its conditional posterior,
\begin{equation}
  \bmath{\xi_i}|\ldots \sim \normal_p\left(\bmath{\hat{\xi}_0^{(i)}}, \bmath{\hat{T}_0^{(i)}}\right),
\end{equation}
where
\begin{eqnarray}
  \bmath{\hat{T}_0^{(i)}} &=& \left[ \left(\bmath{\hat{T}_1^{(i)}}\right)^{-1} + \bmath{T}^{-1} \right]^{-1}\\
  \bmath{\hat{\xi}_0^{(i)}} &=&  \bmath{\hat{T}_0^{(i)}} \left( \bmath{\hat{\xi}_1^{(i)}} + \bmath{T}^{-1}\bmath{\mu} \right). \nonumber
\end{eqnarray}

Once the procedure above is completed, new covariate vectors are drawn for each cluster ($k=1,2,\ldots,K$) given the set of data points residing in it,
\begin{eqnarray}
  \bmath{\xi'_k}|\ldots &\sim& \normal_p(\bmath{\hat{\xi}_2}, \bmath{\hat{T}_2}), \\
  \bmath{\hat{T}_2} &=& \left\{ \bmath{T}^{-1}  + \sum_{i:G_i=k}\left[(\bmath{M_i}^{-1})_{xx} + \transpose{\bmath{\beta}}\bmath{\Sigma}^{-1}\bmath{\beta}\right] \right\}^{-1}, \nonumber\\
  \bmath{\hat{\xi}_2} &=&  \bmath{\hat{T}_2} \Bigg\{  \bmath{T}^{-1}\bmath{\mu} \nonumber\\
  && + \sum_{i:G_i=k}\left[ \left(\bmath{M_i}^{-1}\bmath{z_i}\right)_x  + \transpose{\bmath{\beta}}\bmath{\Sigma}^{-1}\left(\bmath{\eta_i}-\bmath{\alpha}\right) \right] \Bigg\}, \nonumber
\end{eqnarray}
and each new value $\bmath{\xi'_k}$ is assigned to all $\bmath{\xi_i}$ in the corresponding cluster (i.e.\ with $G_i=k$).

\subsubsection{Updating the Dirichlet process concentration}

The procedure for Gibbs sampling $\kappa$ is given by \citet{Escobar2291069}. First, a latent variable, $h$, is introduced and sampled according to a Beta distribution,
\begin{equation}
  h|\ldots \sim \mathrm{Beta}(\kappa+1,\, n).
\end{equation}
Then, $\kappa$ is updated according to
\begin{eqnarray}
  \kappa|\ldots &\sim& \delta\, \mathrm{Gamma}\left[a+K,\, b-\ln(h)\right] \\ 
  && + (1-\delta) \mathrm{Gamma}\left[a+K-1,\, b-\ln(h)\right], \nonumber
\end{eqnarray}
where $a$ and $b$ are the shape and rate parameters of the Gamma prior on $\kappa$, and
\begin{equation}
  \delta = \left[ 1 + n\, \frac{b-\ln(h)}{a+K-1} \right]^{-1}.
\end{equation}
In {\sc lrgs}, the default values of $a$ and $b$ are chosen to be uninformative based on the number of data points, following the prescription given by \citet{Dorazio20093384}.

\subsubsection{Updating the base distribution hyperparameters}

Using the notation of \secref~\ref{sec:dpcovariates}, the hyperparameters of the base distribution can be updated in turn as
\begin{eqnarray}
  \bmath{\mu}|\ldots &\sim& \normal_p\left(\frac{1}{K}\sum_{k=1}^{K}\bmath{\xi'_k},\, \frac{1}{K}\bmath{T}\right), \\
  \bmath{T}|\ldots &\sim& \wishart^{-1}\left[ \sum_{k=1}^{K}(\bmath{\xi'_k}-\bmath{\mu})\transpose{(\bmath{\xi'_k}-\bmath{\mu})} ,\, K+p \right]. \nonumber
\end{eqnarray}

\section{Examples} \label{sec:examples}

This section provides two example applications of the methods discussed above, respectively on a toy model and an astrophysical data set.

\subsection{Toy model} \label{sec:toy}

\begin{figure*}
  \centering
  \includegraphics[scale=0.75]{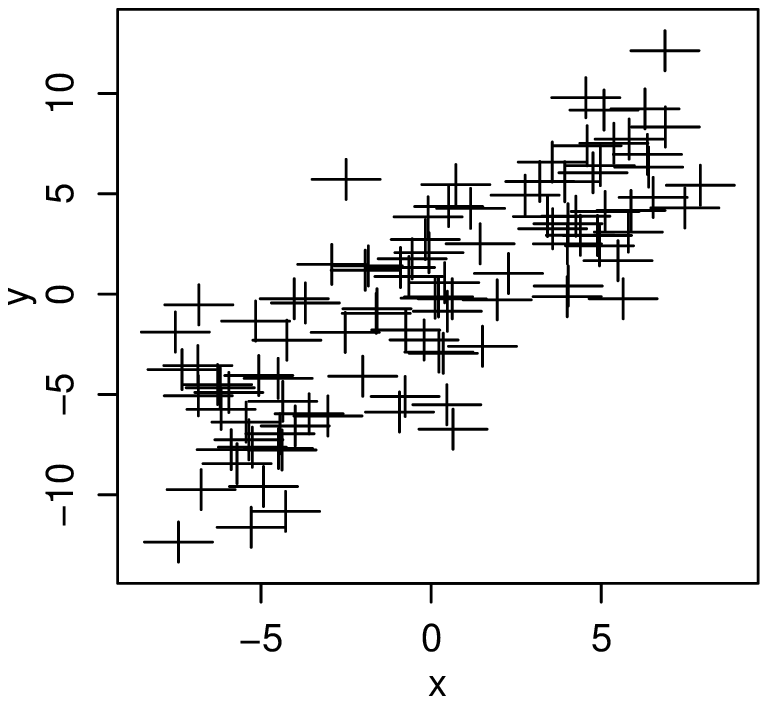}
  \includegraphics[scale=0.75]{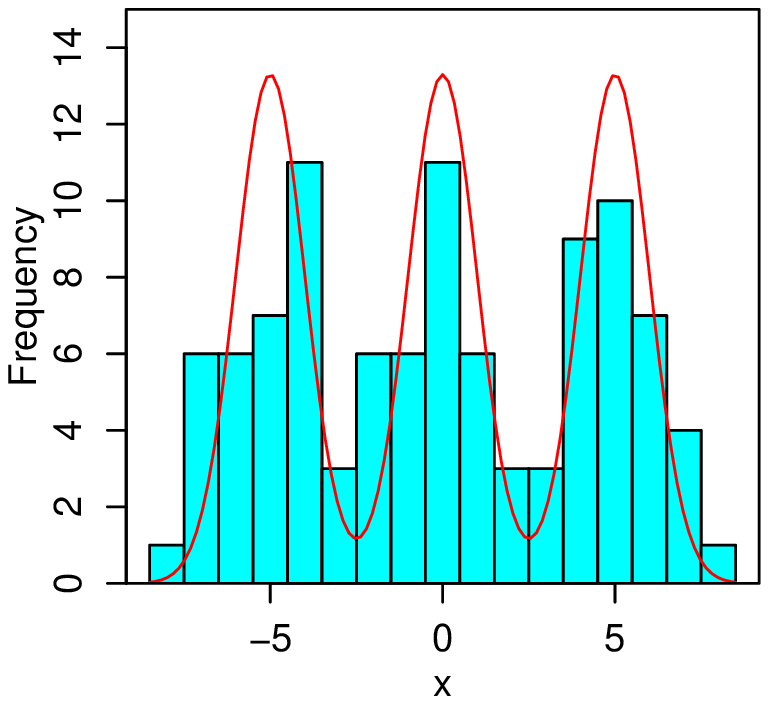}
  \includegraphics[scale=0.75]{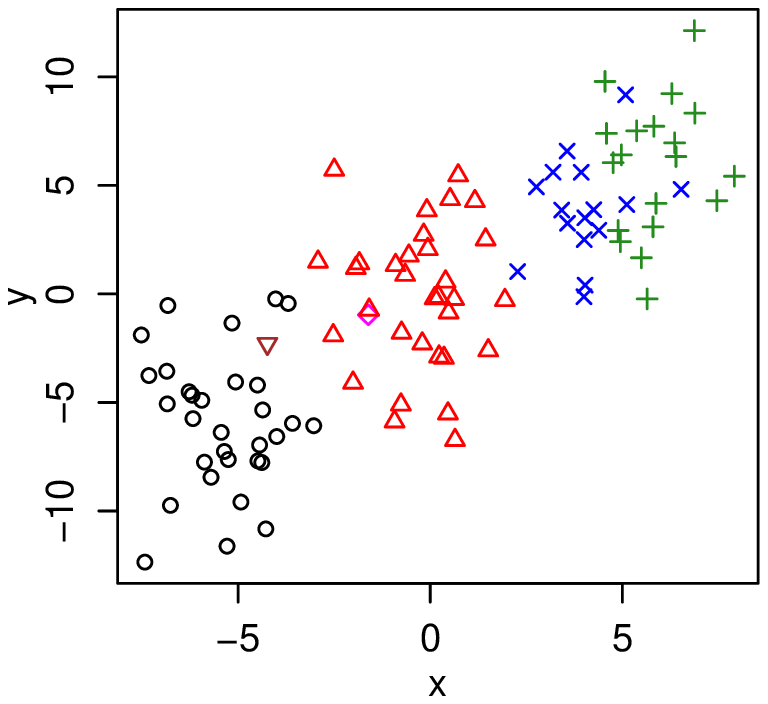}
 \caption[]{
   Left: the simulated data used to fit the toy model in \secref~\ref{sec:toy}. Center: histogram of the simulated covariates, along with the 3-Gaussian-mixture distribution from which they are drawn. Right: the simulated data, with colors/symbols reflecting the cluster assignments of the Dirichlet model at one step in the fit.
  }
  \label{fig:toydata}
\end{figure*}

Consider the case of a single covariate, generated by three distinct Gaussian components, and a single response variable. \tabref~\ref{tab:toy} shows the specific model parameters used to generate the data, and the toy data set is shown in the left panel of \figref~\ref{fig:toydata}. Because the Gaussians generating the covariates are not especially well separated compared to their widths, the presence of three populations is not striking, although a histogram of the measured covariates is suggestive of the underlying structure (center panel of \figref~\ref{fig:toydata}).

\begin{table}
  \centering
  \caption[]{Model parameters used to generate the toy data set in \secref~\ref{sec:toy}. The distribution of covariates is taken to be a mixture of 3 Gaussians.}
  \vspace{1ex}
  \begin{tabular}{cc}
    Parameter & Value \\
    \hline
    $n,p,m,K$ & $100,1,1,3$ \\
    all $\bmath{M_i}$ & $\one_2$ \\
    $\alpha$ & 0 \\
    $\beta$ & 1 \\
    $\Sigma$ & 9 \\
    all $\pi_k$ & $1/3$ \\
    $\mu_k$ & $5(k-2)$ \\
    all $T_k$ & 1 \\
    \hline
  \end{tabular}
  \label{tab:toy}
\end{table}

Suppose we had a physical basis for a 3-component model (or suspected 3 components, by inspection), but wanted to allow for the possibility of more or less structure than a strict Gaussian mixture provides. The Dirichlet process supplies a way to do this. For a given $\kappa$ and $n$, the distribution of $K$ is known,\footnote{Specifically, $K|n,\kappa \sim s(n,K)\,\kappa\,\Gamma(\kappa)/\Gamma(\kappa+n)$, where $s$ is an unsigned Stirling number of the first kind \citep{Antoniak2958336}.} so in principle a prior expectation for the number of clusters, say $3\pm1$,  can be roughly translated into a Gamma prior on $\kappa$. Here I instead adopt an uninformative prior on $\kappa$ \citep{Dorazio20093384}, and compare the results to those of a Gaussian mixture model with $K=3$.

\begin{figure*}
  \centering
  \includegraphics[scale=0.75]{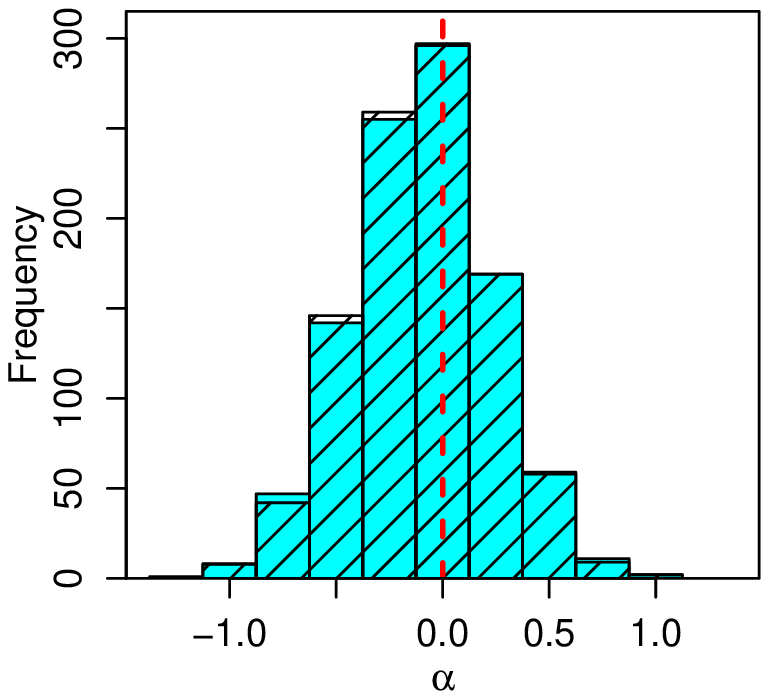}
  \includegraphics[scale=0.75]{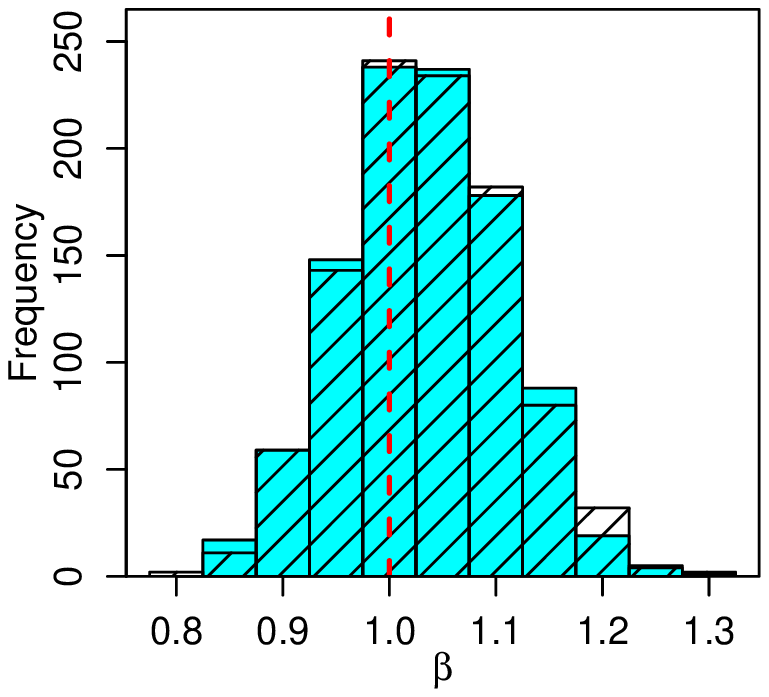}
  \includegraphics[scale=0.75]{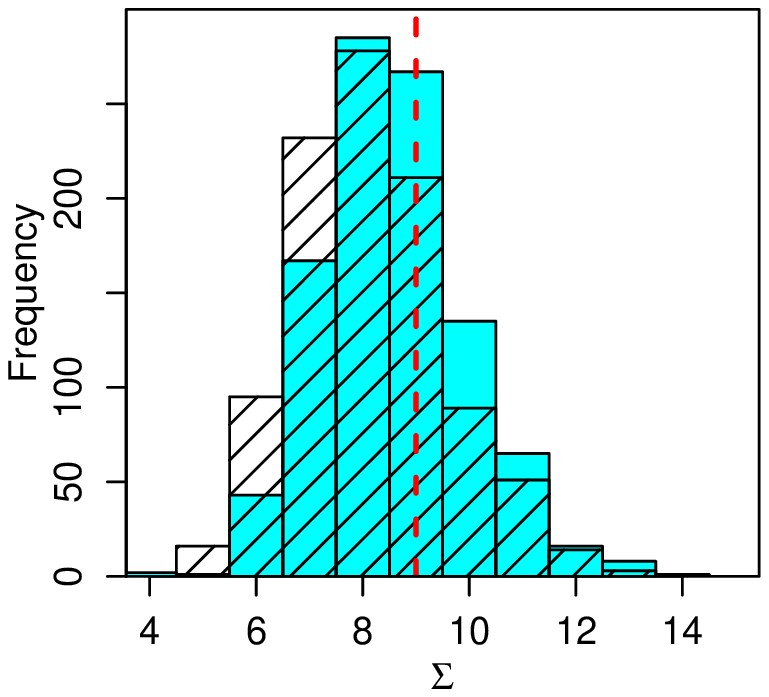}\vspace{3mm}\\
  \includegraphics[scale=0.75]{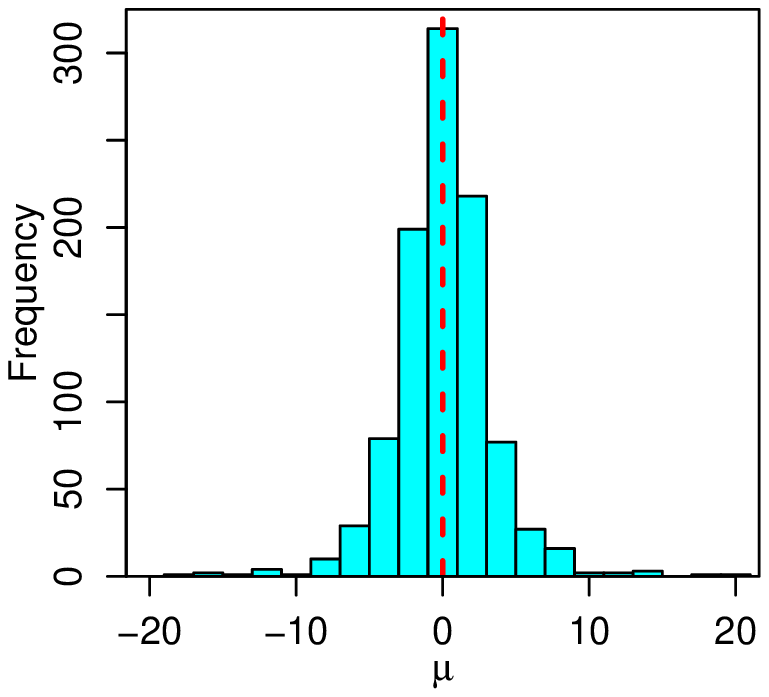}\hspace{5mm}
  \includegraphics[scale=0.75]{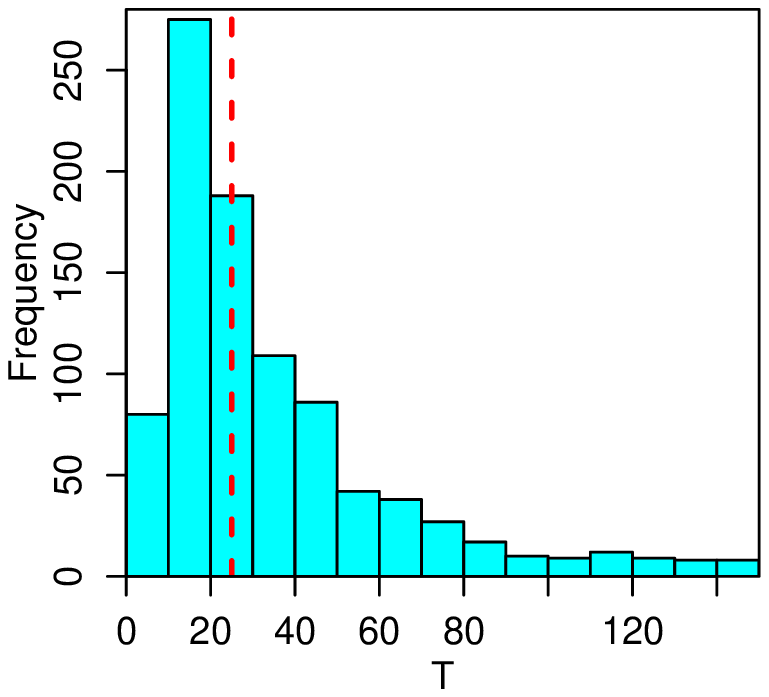}\vspace{3mm}\\
  \includegraphics[scale=0.75]{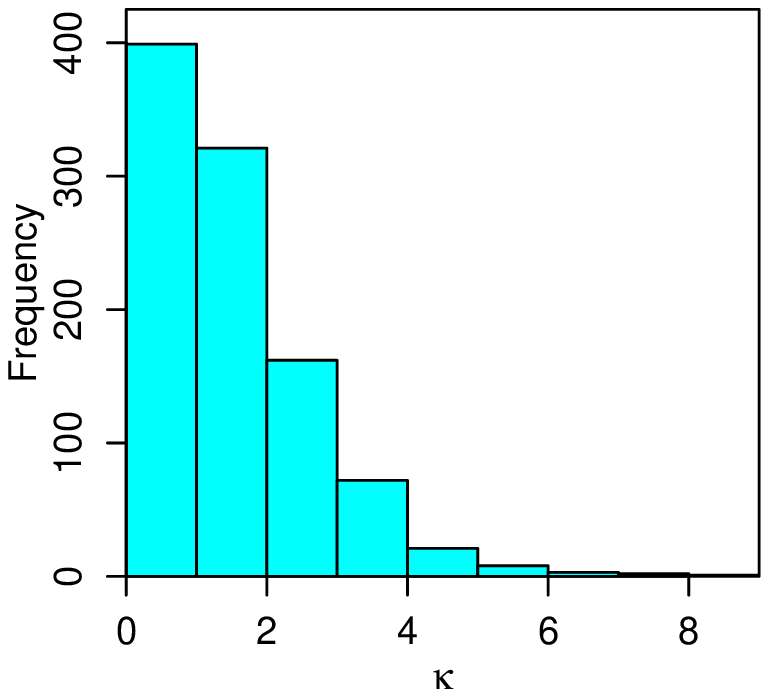}\hspace{5mm}
  \includegraphics[scale=0.75]{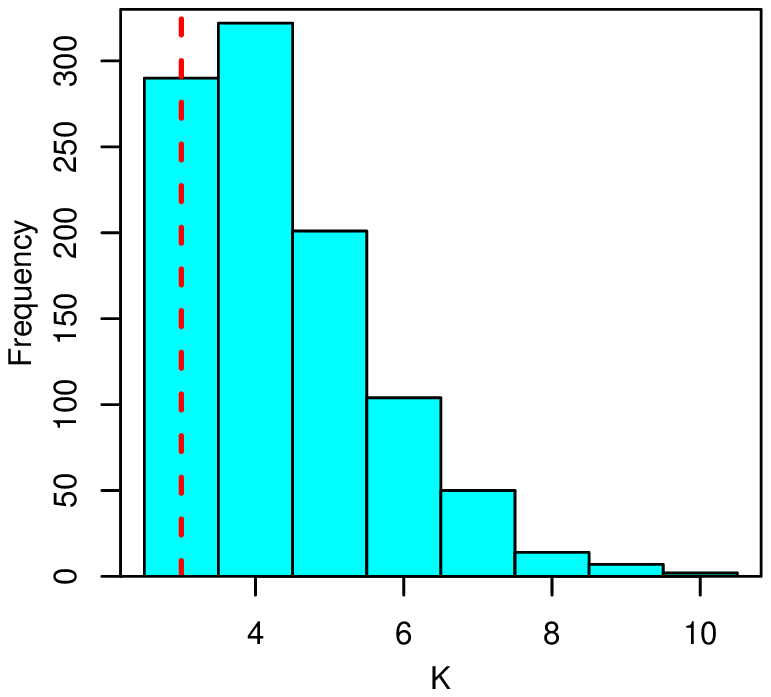}
  \caption[]{
    Histograms of parameter samples using a Dirichlet process (blue shaded) or 3-Gaussian mixture (hatched) prior on the distribution of covariates, for the toy-model analysis of \secref~\ref{sec:toy}. Dashed vertical lines indicate the input values used to generate the data set. The hyperparameters of the Gaussian mixture model are not shown; these correctly converge to the mean and width of the three mixture components.
  }
  \label{fig:toyres}
\end{figure*}

Using the Dirichlet process model, results from a chain of 1000 Gibbs samples (discarding the first 10) are shown as shaded histograms in \figref~\ref{fig:toydata}.\footnote{For this particularly simple problem, the chain converges to the target distribution almost immediately. Comparing the first and second halves of the chain (or multiple independent chains) the Gelman-Rubin $R$ statistic is $<1.01$ for every parameter. The autocorrelation length is also very short, $\ltsim10$ steps for every parameter.} The results are consistent with the input model values (vertical, dashed lines) for the parameters of interest ($\alpha$, $\beta$ and $\Sigma$). The latent parameters describing the base distribution of the Dirichlet process are also consistent with the toy model, although they are poorly constrained. The right panel of \figref~\ref{fig:toydata} shows the cluster assignments for a sample with $K=6$ (the median of the chain); the clustered nature of the data is recognized, although the number of clusters tends to exceed the number of components in the input model.\footnote{We should generically expect this, since it is entirely possible for a mixture of many Gaussians to closely resemble a mixture with fewer components; e.g., in \figref~\ref{fig:toyres}, we see that 2 of the 6 clusters are populated by single data points that are not outliers. The reverse is not true, and here it is interesting that the Dirichlet process cannot fit the data using fewer than $K=3$ clusters (\figref~\ref{fig:toyres}).} An equivalent analysis using a mixture of 3 Gaussians rather than a Dirichlet process model produces very similar constraints on the parameters of interest (hatched histograms in \figref~\ref{fig:toyres}).

\subsection{Scaling relations of relaxed galaxy clusters} \label{sec:astro}

As a real-life astrophysical example, I consider the scaling relations of dynamically relaxed galaxy clusters, using measurements presented by \citet{Mantz1509.01322}. Note that there are a number of subtleties in the interpretation of these results that will be discussed elsewhere; here the problem is considered only as an application of the method presented in this work.

\begin{figure*}
  \centering
  \includegraphics[scale=0.75]{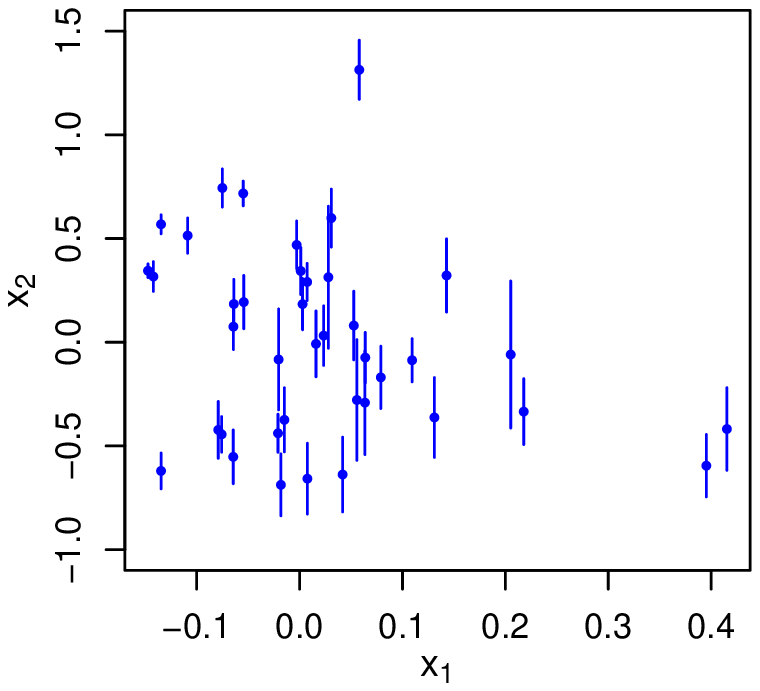}\hspace{5mm}
  \includegraphics[scale=0.75]{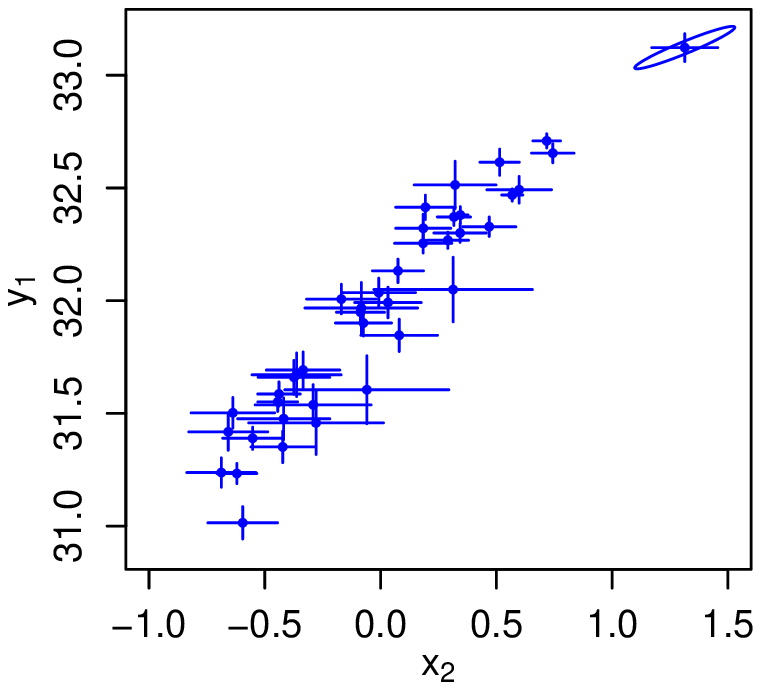}\vspace{3mm}\\
  \includegraphics[scale=0.75]{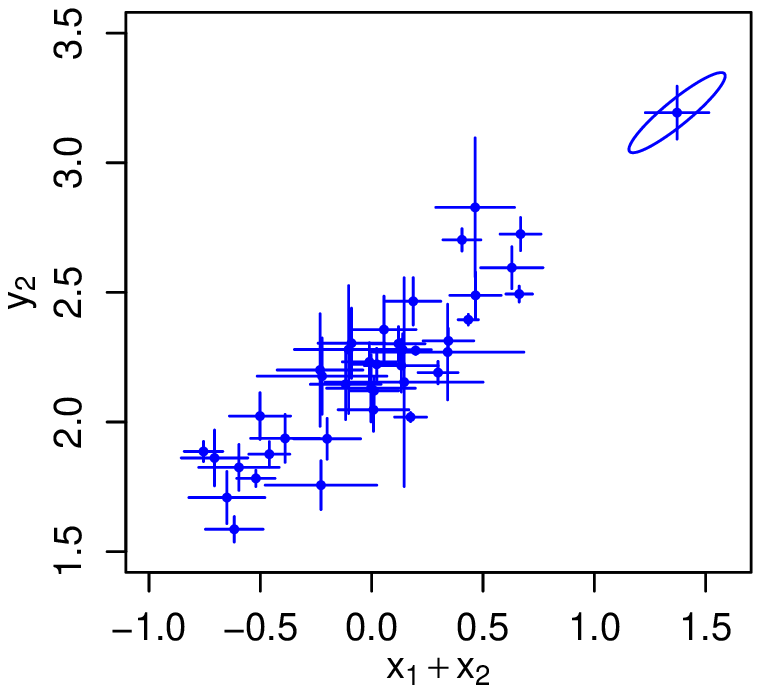}\hspace{5mm}
  \includegraphics[scale=0.75]{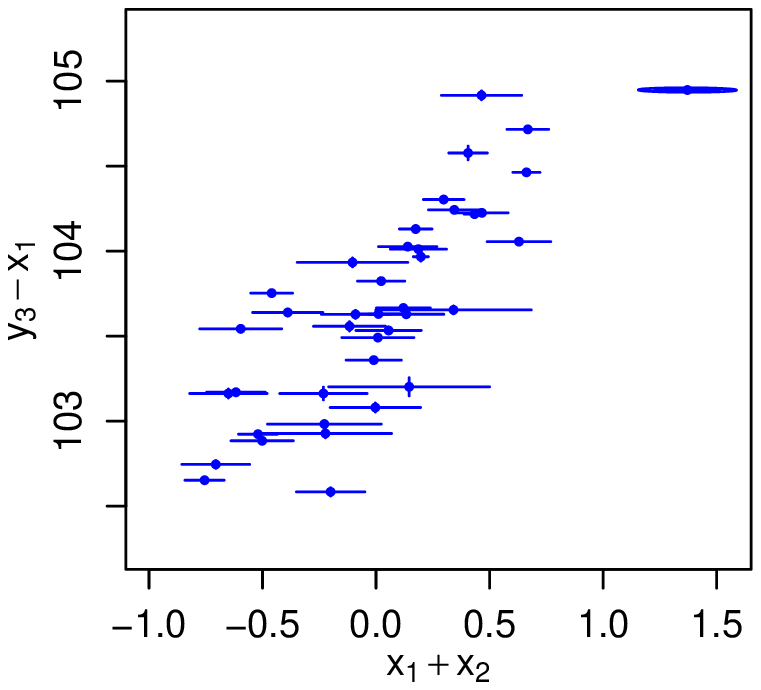}
  \caption[]{
   Scatter plots showing the distribution of measured covariates and responses for the $p=2$, $m=3$ problem of fitting galaxy cluster scaling relations described in \secref~\ref{sec:astro}. An ellipse illustrates the measurement covariance for the most massive (largest $x_2$) cluster in each panel. The particular combinations of $x$ and $y$ plotted are conventional (cf.\ \eqnref~\ref{eq:selfsim}).
  }
  \label{fig:cldata}
\end{figure*}

Briefly, the data set comprises X-ray measurements of 40 massive, relaxed clusters.\footnote{\emph{Galaxy} clusters, not to be confused with the clusters of data points arising in the sampling of the Dirichlet process.} The X-ray observables are total mass, $M$; gas mass, \Mgas{}; average gas temperature, $kT$; and luminosity, $L$. In addition, spectroscopically measured redshifts are available for each cluster. A simple model of cluster formation by spherical collapse under gravity, neglecting gas physics, predicts self-similar power-law scaling relations among these quantities:\footnote{Here I take $L$ to be measured in a soft X-ray band, in practice 0.1--2.4\,keV. Since the emissivity in this band is weakly dependent on temperature for hot clusters such as those in the data set, the resulting scaling relation has a shallower dependence on mass than the more familiar bolometric luminosity--mass relation, $L_\mathrm{bol} \propto E(z) \left[E(z)M\right]^{4/3}$. The exponents in the $L$ scaling of Equation~\ref{eq:selfsim} are specific to the chosen energy band.}
\begin{eqnarray} \label{eq:selfsim}
  \Mgas &\propto& M, \\
  kT &\propto& \left[E(z)\,M\right]^{2/3}, \nonumber\\
  L &\propto& E(z)^{1.92}\,M^{0.92}, \nonumber
\end{eqnarray}
where $E(z)=H(z)/H_0$ is the normalized Hubble parameter at the cluster's redshift. The aim of this analysis is to test whether the power-law slopes above are accurate, and to characterize the joint intrinsic scatter of $\Mgas$, $kT$ and $L$ at fixed $M$ and $z$. Taking the logarithm of these physical quantities, and assuming log-normal measurement errors and intrinsic scatter, this becomes a linear regression with $p=2$ and $m=3$. For brevity, and neglecting units, $(\ln E, \ln M) \rightarrow (x_1, x_2)$ and $(\ln \Mgas, \ln kT, \ln L) \rightarrow (y_1, y_2, y_3)$; I also approximately center the covariates for convenience. \figref~\ref{fig:cldata} shows summary plots of these data. Although measurement errors are shown as orthogonal bars for clarity, the analysis will use a full $5\times5$ covariance matrix accounting for the interdependence of the X-ray measurements (this covariance is illustrated for one cluster in the figure).

Because the redshifts are measured with very small uncertainties, this problem is not well suited to the Dirichlet process prior; intuitively, the number of clusters in the Dirichlet process must approach the number of data points because the data are strongly inconsistent with one another (i.e.\ are not exchangeable). Instead, I use a Gaussian mixture prior with $K=3$, and verify that in practice the results are not sensitive to $K$ (the parameters of interest differ negligibly from an analysis with $K=1$).

\begin{figure*}
  \centering
  \includegraphics[scale=0.75]{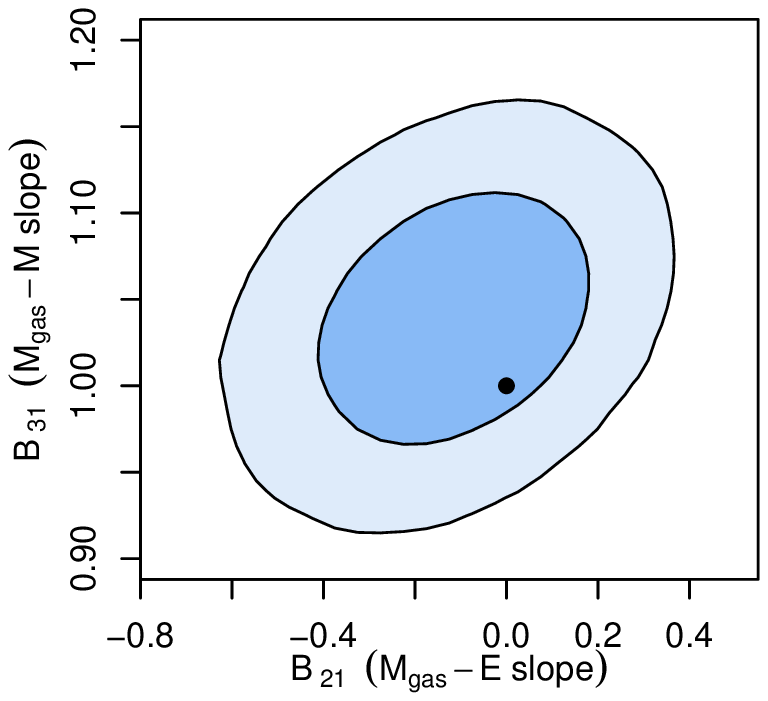}
  \includegraphics[scale=0.75]{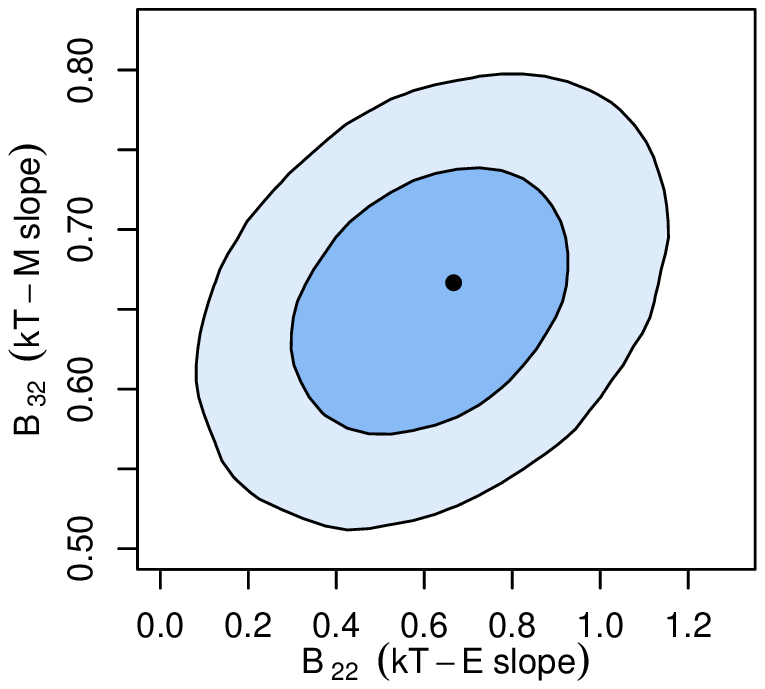}
  \includegraphics[scale=0.75]{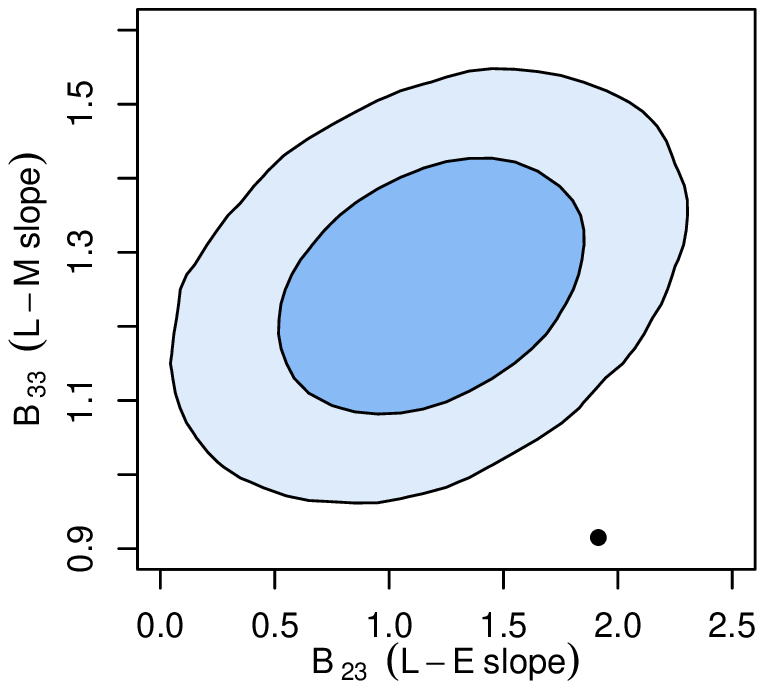}\vspace{3mm}\\
  \includegraphics[scale=0.75]{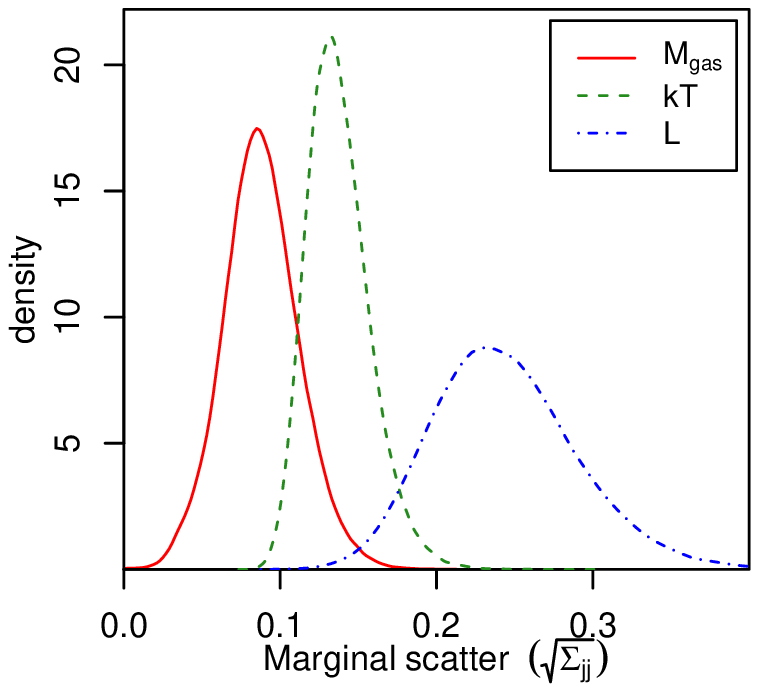}
  \hspace{1cm}
  \includegraphics[scale=0.75]{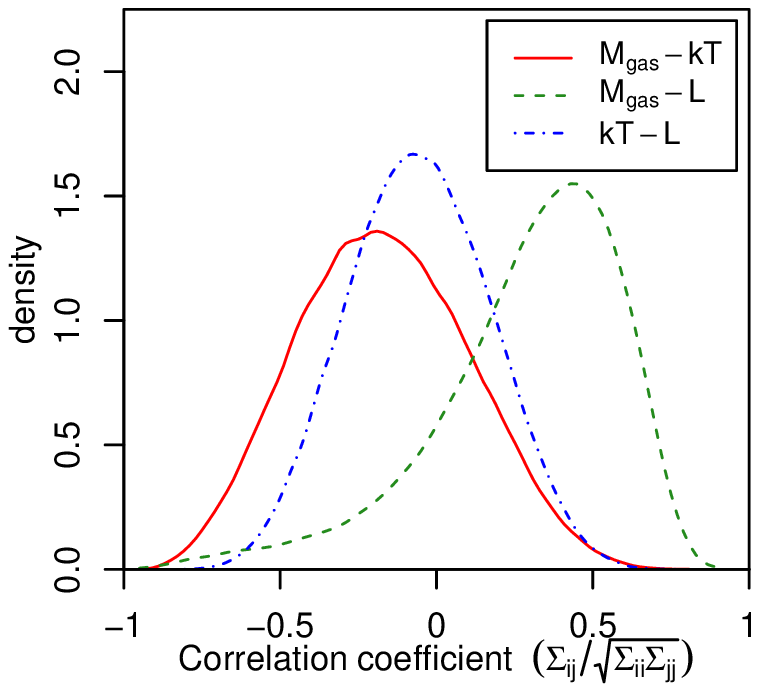}
  \caption[]{
    Top row: joint 68.3 and 95.4 per cent confidence regions on the slope parameters of each of the scaling relations in \eqnref~\ref{eq:selfsim}. Black circles indicate the self-similar expectation given in \eqnref~\ref{eq:selfsim}. Bottom left: posterior distributions for the marginal intrinsic scatter parameters of the model. Bottom right: posteriors for the off-diagonal intrinsic scatter terms, expressed as correlation coefficients.
  }
  \label{fig:clres}
\end{figure*}

Marginalized 2-dimensional constraints on the power-law slopes of each scaling relation are shown in the top row of \figref~\ref{fig:clres} (68.3 and 95.4 per cent confidence). On inspection, only the luminosity scaling relation appears to be in any tension with the expectation in \eqnref~\ref{eq:selfsim}, having a preference for a weaker dependence on $E(z)$ and a stronger dependence on $M$. These conclusions are in good agreement with a variety of earlier work (e.g.\ \citealt{Reiprich0111285, Zhang0702739, Zhang0802.0770, Mantz0909.3099, Rykoff0802.1069, Pratt0809.3784, Vikhlinin0805.2207, Leauthaud0910.5219, Reichert1109.3708, Sereno1502.05413}; see also the review of \citealt{Giodini1305.3286}).

\begin{figure*}
  \centering
  \includegraphics[scale=0.75]{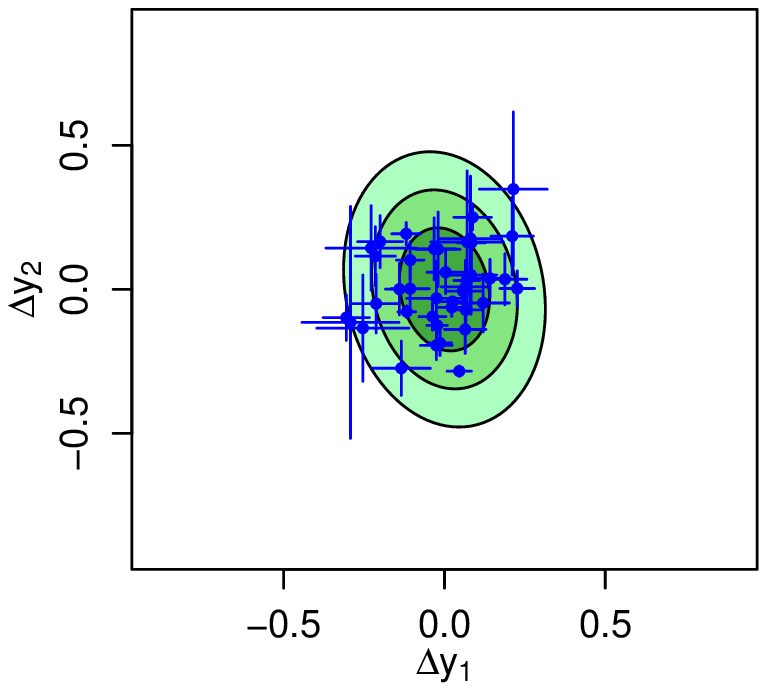}
  \includegraphics[scale=0.75]{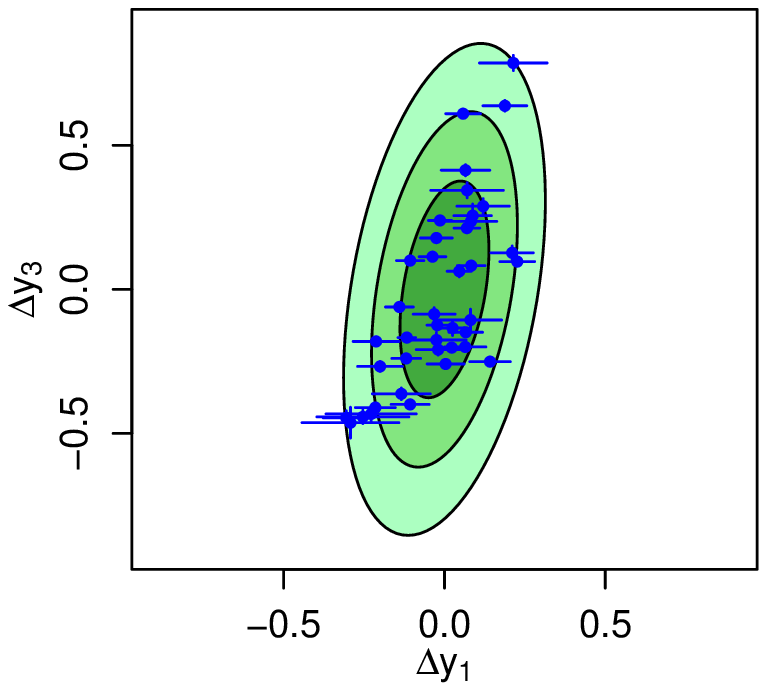}
  \includegraphics[scale=0.75]{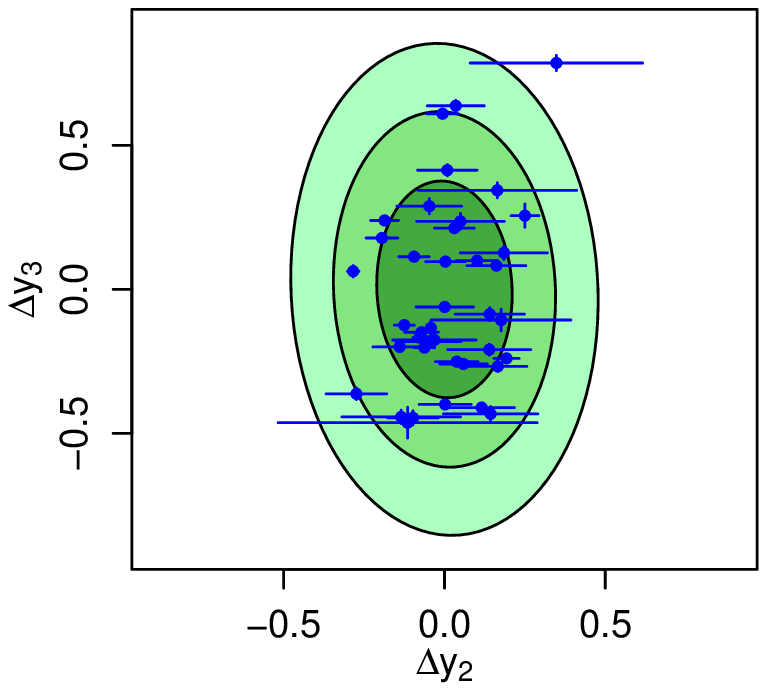}
  \caption[]{
    Residuals of $\bmath{y}$ with respect to the best-fitting values of $\bmath{x}$ and the best-fitting scaling relations. For clarity, measurement errors are shown as orthogonal bars, even though the measurement covariances are non-trivial. In particular, the ``by-eye'' positive correlation in the left panel is due to a positive correlation in the measurement uncertainties. Shaded ellipses correspond to 1, 2 and $3\sigma$ intrinsic scatter (in the 2 dimensions shown in each panel), according to the best-fitting intrinsic covariance matrix.
  }
  \label{fig:clresid}
\end{figure*}

The posterior distributions of the elements of the multi-dimensional intrinsic covariance matrix are shown in the bottom row of \figref~\ref{fig:clres}, after transforming to marginal scatter (square root of the diagonal) and correlation coefficients (for the off-diagonal elements). The intrinsic scatters of $\Mgas$ and $kT$ at fixed $M$ and $z$ are in good agreement with other measurements in the literature (see \citealt{Allen1103.4829, Giodini1305.3286}, and references therein); the scatter of $L$ is lower than the $\sim40$ per cent typically found, likely because this analysis uses a special set of morphologically similar clusters rather than a more representative sample. The correlation coefficients are particularly challenging to measure, and the constraints are relatively poor. Nevertheless, the ability to efficiently place constraints on the full intrinsic covariance matrix is an important feature of this analysis. Within uncertainties, these results agree well with the few previous contraints on these correlation coefficients in the literature \citep{Mantz0909.3099, Maughan1212.0858}. The best-fitting intrinsic covariance matrix is illustrated visually in \figref~\ref{fig:clresid}, which compares it to the residuals of $\bmath{y}$ with respect to the best-fitting values of $\bmath{x}$ and the best-fitting scaling relations.

\section{Summary}

I have generalized the Bayesian linear regression method described by \koseven{} to the case of multiple response variables, and included a Dirichlet process model of the distribution of covariates (equivalent to a Gaussian mixture whose complexity is learned from the data). The algorithm described here is implemented independently of the {\sc linmix\_err} IDL code of \koseven{} as an {\sc r} package called {\sc lrgs}, which is publicly available. Two examples, respectively using a toy data set and real astrophysical data, are presented.

A number of further generalizations are possible. In principle, significant complexity can be added to the model of the intrinsic scatter in the form of a Gaussian mixture or Dirichlet process model (with a Gaussian base distribution) while maintaining conjugacy of the conditional posteriors, and thereby the efficiency of the Gibbs sampler. The case censored data (upper limits on some measured responses) is discussed by \koseven{}. This situation, or, more generally, non-Gaussian measurement errors, can be handled by rejection sampling (at the expense of efficiency) but is not yet implemented in {\sc lrgs}. Also of interest is the case of truncated data, where the selection of the data set depends on one of the response variables, and the data are consequently an incomplete and biased subset of a larger population. This case can in principle be handled by modeling the selection function and imputing the missing data \citep{Gelman2004BayesianDataAnalysis}. {\sc lrgs} is shared publicly on GitHub, and I hope that users who want more functionality will be interested in helping develop the code further.

\section*{Acknowledgments}

The addition of Dirichlet process modeling to this work was inspired by extensive discussions with Michael Schneider and Phil Marshall. Anja von der Linden did some very helpful beta testing. I acknowledge support from the National Science Foundation under grant AST-1140019.

\def \aap {A\&A} 
\def \apj {ApJ}
\def \araa {ARA\&A}
\def \mnras {MNRAS}
\def \ssr {Space Sci.\ Rev.}


\end{document}